\documentstyle[aps]{revtex}
\begin{document}
\title{Breathing and randomly walking pulses in a semilinear Ginzburg-Landau system}
\author{Hidetsugu Sakaguchi$^{a}$ and Boris A. Malomed$^{b}$}
\address{$^{a}$
Department of Applied Science for Electronics and Materials, \\
Interdisciplinary Graduate School of Engineering Sciences, Kyushu University,%
\\
Kasuga, Fukuoka 816-8580, Japan//\\
$^{b}$Department of Interdisciplinary Studies, Faculty of Engineering,\\
Tel Aviv University, Tel Aviv 69978, Israel}
\maketitle

\begin{center}
{\bf Abstract}
\end{center}

We consider a system consisting of the cubic complex Ginzburg-Landau
equation which is linearly coupled to an additional linear equation. The
model is known in the context of dual-core nonlinear optical fibers with one
active and one passive cores. By means of systematic simulations, we find
new types of stable localized excitations, which exist in the system in
addition to the earlier found stationary pulses. The new localized
excitations include pulses existing on top of a small-amplitude background
(that may be regular or chaotic) {\em above} the threshold of instability of
the zero solution, and breathers into which stationary pulses are
transformed through a Hopf bifurcation below the aforementioned threshold. A
sharp border between stable stationary pulses and breathers, which precludes
their coexistence, is identified. Stable bound states of two breathers with
a phase shift $\pi /2$ between their internal vibrations are found too.
Above the threshold, the pulse is standing if the background oscillations
are regular; if the background is chaotic, the pulse is randomly walking.
With the increase of the system's size, additional randomly walking pulses
are spontaneously generated. The random walk of different pulses in a
multi-pulse state is partly synchronized due to their mutual repulsion. At a
large overcriticality, the multi-pulse state goes over into a spatiotemporal
chaos. \newline
{\it PACS:} 05.45; 42.65.-k; 47.52+j; 47.54.+r\newline
{\it Keywords:} Localized pulse, complex Ginzburg-Landau equation, random
walk, breather \newpage 

\section{Introduction}

The important role played by localized pulses (sometimes called
``autosolitons'' \cite{book}) in models of the Ginzburg-Landau (GL) type 
\cite{CH} is well known. They find applications to plasmas and plasma-like
media \cite{book}, \cite{Stenflo}-\cite{Kawahara2}, nonequilibrium
semiconductors \cite{book}, hydrodynamics (Poiseuille flow \cite{Poiseille},
electrohydrodynamic
convection \cite{electro}), nonlinear optics 
\cite{Winful}-\cite{Petersburg2} , etc.

The simplest model which gives rise to localized pulses is the complex cubic
GL equation, 
\begin{equation}
u_{t}=\gamma _{0}u-\left( g-i\sigma \right) |u|^{2}u+(\gamma _{1}+i\gamma
_{2})u_{xx},  \label{cubic}
\end{equation}
where $\gamma _{0}$, $g$, and $\gamma _{1}$ are positive coefficients of the
linear gain, nonlinear losses, and effective diffusion (dispersive losses),
while $\sigma $ and $\gamma _{2}$ (which may have either sign) account for
the nonlinear frequency shift and dispersion, respectively. Equation (\ref
{cubic}) always has a single exact solitary-pulse (SP) solution \cite
{Poiseille,Stenflo}, which is, however, unstable, as its background, i.e.,
the trivial solution $u=0$, is obviously unstable due to the presence of the
linear gain.

The simplest possibility to modify the model so that to produce {\em stable}
SPs is to convert it into the quintic GL equation \cite{quintic}, setting $%
\gamma _{0}<0$ and $g<0$, and adding a nonlinear quintic dissipative term in
order to provide for the overall stabilization. In the quintic GL equation,
the trivial solution is stable since $\gamma _{0}<0$, and two solitary-pulse
solutions, one unstable and one stable, may coexist at fixed values of
parameters \cite{quintic2}.

In this connection, it is necessary to mention a modification of the usual
cubic GL equation proposed by Sch\"{o}pf and Kramer \cite{Kramer} (see also
the works \cite{others}), in which $\gamma _{0}=1$ and $g$
is negative, while no higher-order stabilizing nonlinearity is
added. The corresponding model lacks global stability, as a very large
perturbation with a small wavenumber will provoke blowup. Nevertheless,
numerical simulations have shown that finite-amplitude solutions in the form
of both chaotic and stationary arrays of pulses may exist in this model, due
to the stabilizing effect exerted by the dispersion and diffusion.

The quintic GL equation is, generally, a phenomenological model, as the
derivation from the first principles can scarcely stop at the fifth-order
nonlinearity (provided that it is a crucially important term rather than a
small correction). Although this model can sometimes describe experimentally
observed waves with a fairly high accuracy, an example being the so-called
dispersive chaos in the traveling-wave convection in binary fluids \cite
{Kurths}, it is desirable to find more models that allow for the existence
of stable SPs and can be derived from the first principles. A possibility is
to consider a model of a dual-core nonlinear optical fiber, in which the
linear gain, dispersion, effective diffusion (spectral filtering), and Kerr
(nondissipative cubic) nonlinearity are present in one (active) core, while
the other (passive) one, linearly coupled to the active core, has only
linear loss \cite{Winful,Atai1}. Thus, the model consists of a cubic GL
equation linearly coupled to the second, purely linear, ordinary
differential equation for the local amplitudes $u$ and $v$ of the waves in
the two subsystems: 
\begin{eqnarray}
u_{t} &=&\gamma _{0}u+i\sigma |u|^{2}u+(\gamma _{1}+i\gamma _{2})u_{xx}+iv, 
\nonumber \\
v_{t} &=&-\Gamma v+iu  \label{model}
\end{eqnarray}
(note that, in the application to optical fibers, the evolutional variable $t
$ is actually the distance along the fiber, while $x$ is the so-called local
time \cite{optics}). The parameters $\gamma _{0}$, $\gamma _{1}$, $\gamma
_{2}$ and $\sigma $ of the active subsystem have the same meaning as in the
cubic GL equation (\ref{cubic}), there is no saturating nonlinear term, $\Gamma $ is the loss factor in the passive
subsystem, and the coefficient of the linear coupling between the subsystems
is set equal to $1$. 

Note that an essentially more general version of the model, with the
dispersion, Kerr, and filtering terms present in the equation for the
passive core was also studied in detail \cite{Winful,Atai2}. However, both
physical arguments and numerical results presented in \cite{Atai1} clearly
demonstrate that the version of the model with the very simple form of the
additional equation displayed above is quite sufficient to grasp all the
essential dynamical features of the dual-core optical fiber with one active
and one passive cores \cite{Atai1}. 

Although the model (\ref{model}), as well as those considered in refs. \cite
{Kramer} and \cite{others}, has only cubic nonlinearity, it is very
different from them, having another stabilization mechanism. In particular,
it has been demonstrated analytically and numerically that there is no
global instability in the present model (i.e., it never gives rise to
blowup) \cite{Winful,Atai1}.

Parameters in eqs. (\ref{model}) can be chosen so that to provide for the
stability of the zero solution, which lends localized pulses, if any, a
chance to be stable \cite{Winful,Atai1,Atai2,new}, see below. Then, two {\em %
exact} zero-velocity SP solutions to eqs. (\ref{model}) can be found,
following the pattern of the exact solution to the cubic GL equation, in the
form
\begin{equation}
u=u_{0}[{\rm sech}(kx)]^{(1+i\mu )}e^{i\omega t},\;v=v_{0}[{\rm sech}%
(kx)]^{(1+i\mu )}e^{i\omega t}  \label{exactSP}
\end{equation}
\cite{Atai1}. The two exact solutions have different values of $u_0,v_0,k,\mu$ and $\omega$. Direct numerical simulations have clearly shown that one of
the two exact SPs may be stable \cite{Atai1,Atai2}. The simulations have
revealed nontrivial stability borders for this SP in the parameter space of
the model (``nontrivial'' implies that the stability borders are different
from stability conditions for the zero background). Interactions between
stable SPs were also simulated in detail in this model \cite{Atai2}; in
particular, it has been found that stable bound states of two pulses exist.
Three-pulse states exist too, but they are unstable against
symmetry-breaking perturbations.

The objective of this work is to search for stable localized states in the
model (\ref{model}) beyond instability borders of the stationary pulses. In
section 2, we demonstrate that effectively stable pulses and sets of pulses
exist in the region where the zero background is {\it unstable}. It should
be mentioned that a possibility for a pulse to survive above the
background-instability threshold is known in a GL model of another type \cite
{Dutch}, but in that case the stabilization is provided by the fact that the
pulse is moving, due an extra symmetry-breaking term added to the GL
equation, and, in a system with periodic boundary conditions, it may perform
a round trip quickly enough to suppress the growing local perturbations. In
our model, the situation is different: immediately above the
background-instability threshold, we observe stable pulses resting on top of
a background standing-wave pattern with a small amplitude. With further
increase of the overcriticality, the background pattern keeps a rather small
amplitude but becomes chaotic. In the course of this transition, SPs remain
stable, although they become {\em randomly walking} ones, rather than
keeping zero velocity. While we did not try to elaborate in detail how the
randomization of the background gives rise to the random walk of the pulse,
the effect seems to be quite similar to the Brownian motion, resembling the 
{\it Gordon-Haus effect} (random walk of solitons interacting with random
noise) in optical fibers \cite{optics}. Moreover, additional pulses are
spontaneously generated at larger values of the overcriticality, and
spectacular correlations in the random walk of far separated pulses are
observed in this case, due to their mutual repulsion.

Another finding, presented in section 3, is that one of generic modes of the
instability of stationary SPs (in the case when the zero background is
stable) does not destroy them, but instead transforms them into stable {\it %
breathers} (vibrating SPs). Stable bound states of two breathers were found
too, while bound states with a larger number of breathers were not found.
All these types of stable nonstationary SPs are fairly novel to the present
model, and may be of considerable interest for the general analysis of
models of the GL type.

\section{Solitary pulses above the background-instability threshold}

\subsection{Stability of the zero background}

As it was explained above, the first necessary condition for the unequivocal
stability of SPs is stability of the zero solution. To investigate this
issue, one should linearize eqs. (\ref{model}) and look for a perturbation
eigenmode $\sim \exp \left( \lambda t+iqx\right) $, where $q$ is an
arbitrary real perturbation wavenumber, and $\lambda $ is the corresponding
instability growth rate. This yields a dispersion equation for $\lambda (q)$ 
\cite{Winful,Atai1,Atai2,new}: 
\begin{equation}
\lambda ^{2}+(\Gamma -\gamma _{0}+\gamma _{1}q^{2}+i\gamma _{2}q^{2})\lambda
+\Gamma (-\gamma _{0}+\gamma _{1}q^{2}+i\gamma _{2}q^{2})=0\,.
\label{lambda}
\end{equation}
Onset of instability takes place when ${\rm Re\,}\lambda $ changes its sign
at some (critical) value of $q$. As it follows from eq. (\ref{lambda}), this
critical condition gives rise to an equation for the critical value: 
\begin{equation}
\frac{\gamma _{2}^{2}q^{4}\Gamma (-\gamma _{0}+\gamma _{1}q^{2})}{(\Gamma
-\gamma _{0}+\gamma _{1}q^{2})^{2}}+\Gamma (-\gamma _{0}+\gamma
_{1}q^{2})+1=0.  \label{critical}
\end{equation}

In the further analysis, we will display typical numerical results obtained
at fixed values 
\begin{equation}
\sigma =1,\gamma _{2}=5,\gamma _{1}=0.9\,,  \label{fixed}
\end{equation}
while $\gamma _{0}$ will be varied at some fixed values of $\Gamma $. Note
that the variation of the gain parameter $\gamma _{0}$ is the most
physically meaningful way to scan the results, as it can be easily changed
in the optical experiment by adjusting the pump power, while the other
parameters are fixed for a given experimental setup.

Figure 1 displays the line $\gamma _{0}(q)$, as obtained from eq. (\ref
{critical}) ({\it neutral stability curve}) at the fixed values (\ref{fixed}%
) and $\Gamma =1.35$. Above the line, one has ${\rm Re}\,\lambda (q)>0$ with 
${\rm Im}\,\lambda (q)\neq 0$, i.e., oscillatory instability takes place.
The curve has a minimum $\left( \gamma _{0}\right) _{\min }=0.527$ at $q=0.54
$. That is, the zero solution is stable if $\gamma _{0}\leq \left( \gamma
_{0}\right) _{\min }$, and when the critical gain $\left( \gamma _{0}\right)
_{\min }$ is reached, the\ zero background becomes unstable against
finite-wavenumber perturbations. In view of the importance of the critical
value of the gain, in fig. 2 we display it as a function of $\Gamma $. The
cusp at $\Gamma =1.82$ is related to a switch of the instability mode at
this point: at $\Gamma <1.82$, the finite-wavenumber instability takes
place, while at $\Gamma >1.82$ the value $\gamma _{0}=\left( \gamma
_{0}\right) _{\min }$ is attained at $q=0$.

\subsection{Standing and walking solitary pulses}

Simulations of the dynamics of SPs were performed by means of a
pseudospectral method, assuming periodic boundary condition in $x$, with the
period $60$ or $120$ (in some cases, the period was $480$, see below). As
the initial configuration, we  took the exact SP solutions (\ref
{exactSP}), borrowing expressions for its parameters from \cite{Atai1}. In
those cases when SPs were found to be stable in the simulations reported in
Ref. \cite{Atai1}, we also saw that they were stable. In this section, we
focus on a possibility of existence of robust pulses {\em above} the
critical gain $\left( \gamma _{0}\right) _{\min }$, when it appears that SPs
cannot be stable.

Figure 3 displays SP produced by the simulations at $\gamma _{0}=0.54$ and $%
\Gamma =1.35$. As it is seen from fig. 1, this point is located slightly
above the neutral stability curve, so that the corresponding overcriticality
parameter is $\epsilon \equiv \left[ \gamma _{0}-\left( \gamma _{0}\right)
_{\min }\right] /\left( \gamma _{0}\right) _{\min }\approx 0.025$. As is
seen in fig. 3a, the weak background instability generates a small-amplitude
standing-wave pattern, the stable localized pulse existing on top of it. To
further illustrate the properties of this solution, in fig. 3b we specially
display the time dependence of ${\rm Re\,}u$ at two points: in the center of
the localized pulse, and in the background. Both dependences are periodic,
but their frequencies are very different (the background oscillations are
much slower), i.e., SP and the background are {\em not} synchronized.

The background oscillations are regular at $\gamma _{0}=0.54$; however, they
become {\em chaotic} for $\gamma _{0}>0.55$, at the same value $\Gamma =1.35$%
. We use the term "chaotic", since  the time evolution of the background 
is irregular and has no periodicity.
Figure 4a displays a localized pulse existing on top of the chaotic
background at $\gamma _{0}=0.61$, the corresponding overcriticality being $%
\epsilon =0.157$. Figure 4(b) displays the time evolution of $|u(x,t)|$.
Interaction of the localized pulse with the chaotic background apparently
gives rise to a {\em random walk} of the pulse; however, the pulse is not
destroyed. The temporal evolution of the SP's central coordinate $X_{p}$ is
shown in fig. 5a. To analyze the character of the random walk of the
localized pulse, we display in fig. 5(b) the time-averaged squared
displacement $\left\langle \left( \Delta X_{p}\right) ^{2}\right\rangle $,
where $\Delta X_{p}\equiv \Delta X_{p}(t+\Delta t)-\Delta X_{p}(t)$, vs. the
temporal interval $\Delta t$. As it is obvious from fig. 5b, the random walk
approximately obeys the diffusive law, $\left\langle \left( \Delta
X_{p}\right) ^{2}\right\rangle \sim \Delta t$.

{\em Spontaneous formation} of new pulses takes place for larger $\gamma _{0}
$. Note that spontaneous generation of effectively stable pulses by the
unstable background was observed in simulations of another GL model in Ref. 
\cite{Dutch}. It is not quite clear whether there is a definite border for
the spontaneous generation of SPs from the chaotic background. The pattern
shown in fig. 6 was generated from a nearly zero initial state (transient
evolution is not shown) at $\gamma _{0}=0.64$, corresponding to\ $\epsilon
=0.214$. Three localized pulses are spontaneously generated in this case.
The background is chaotic, and the three SPs are not identical. It is
clearly observed that the pulses repel each other, so that they keep nearly
equal separations between themselves (recall we simulated the system (\ref
{model}) with periodic boundary conditions). Figure 6b displays the time
evolution of $|u(x,t)|$. A noteworthy feature of the established dynamical
state is that the random walk of all the three pulses is highly
synchronized, so that the pulses keep nearly equal separations between
themselves in the course of the random walk. The number of the spontaneously
generated pulses grows with the system's size. In fig. 7 we show the
established state at the same values of the parameters as in fig. 6, but
with the size of the system $480$ instead of $120$. The separations between
SPs remain nearly equal, while the synchronization of the random walk of SPs
is weaker in a large system with the chaotic background. The synchronization
does take place between adjacent pulses, but the full global synchronization
does not occur, as is seen in fig. 7.

The number of the pulses in an established state increases with $\gamma_0$
but it is not a uniquely defined function of the system's parameters and
size. More detailed simulations (not shown here) demonstrate that extra
pulses can be added, and some may be removed, i.e., their exact number
depends on the way the state was made.

Finally, at large values of the overcriticality, many pulses are created,
and the system goes over into a more irregular ``turbulent'' state (in which there is no spatially regular structure of localized pulses).
The transition occurs close to at $\gamma _{0}=1.0$; however, this depends
on the number of pulses. Examples are shown in fig. 8 for $\gamma _{0}=1.05$
(a) and $\gamma _{0}=2$ (b) (again, with $\Gamma =1.35$). Note that these
states have been obtained starting from nearly-zero initial conditions.
Generation of pulse sets and the turbulent behavior are seen in these
pictures. The time evolution is very fast, and the final number of pulses is
larger for $\gamma _{0}=2.$

\section{Breathers and their bound states}

The exact pulse solutions to eqs. (\ref{model}) found in \cite{Atai1} are
stable only in a part of the parameter region where they exist \cite
{Atai1,Atai3}. Besides the loss of the stability of the zero background,
other, more nontrivial destabilization mechanisms occur as well \cite{Atai1}%
. In particular, it seems very plausible that the stationary pulse can be
destabilized through a Hopf bifurcation.

In fig. 9, we show a stable breather replacing the stationary pulse at $%
\gamma _{0}=0.18$ and $\Gamma =0.2$. Figure 9a displays the time evolution
of the profile $|u(x,t)|$, and fig. 9(b) displays the time evolution of $%
|u(x_{p},t)|$ at the peak position. Limit-cycle intrinsic oscillations of
the pulse are clearly seen. Breathing localized pulses were earlier found in
a reaction-diffusion model \cite{Koga} and in the quintic GL equation \cite
{Brand} (note that the breathing solution in fig. 9a has a form similar to
the well-known two-soliton solution of the nonlinear Schr\"{o}dinger
equation,  $u=4\eta \lbrack \cosh (3kx)+3\exp (4i\eta ^{2}t)\cosh
(kx)]/[\cosh (4kx)+4\cosh (2kx)+3\cos (4\eta ^{2}t)]$). To present a whole
family of the breathers, in fig. 10a we show the amplitude of the
oscillations of $|u(x_{p},t)|$ at the breather's central point as a function
of $\gamma _{0}$ for $\Gamma =0.2$. 
At the value $\gamma _{0}\approx 0.156$,
at which the oscillation amplitude vanishes, the breather changes into the
stationary SP.  Figure 10b displays the square amplitude $A^2$ as a function of  $\gamma_0$ near the critical point. The linear dependence of $A^2$ 
on $\gamma_0$ implies $A\propto\sqrt{\gamma_0-\gamma_{0c}}$ near the critical point $\gamma_{0c}$, that is, the bifurcation is the supercritical Hopf bifurcation. 

If the gain parameter $\gamma _{0}$ becomes too small, no nontrivial
solution can exist in the system. Thus, there is a minimum value of $\gamma
_{0}$ confining the existence region of the localized pulses. Collecting
results from many runs of the simulations performed at different values of $%
\gamma _{0}$ and $\Gamma $, while the other parameters took the fixed values
(\ref{fixed}), in fig. 11 we have drawn the phase diagrams of the localized
solutions existing {\em below} the threshold of the zero-background's
instability. The zero background is stable beneath the line marked by
squares, the breathing pulses change into stationary ones at the line marked
by crosses, and the stationary pulses are changed by the zero solution at
the line marked by diamonds. In accord with what was said above, the border
between the stationary pulses and breathers (crosses in fig. 11) is {\em %
sharp}, i.e., there is no overlapping between them. That is a natural consequence of the {\em super}critical character of the
Hopf bifurcation (see above), which makes coexistence of stable stationary
and oscillating solutions very implausible.

The Hopf-bifurcation line collides with the zero-background's instability
line near $\Gamma =0.3$ and $\gamma _{0}=0.27$. For $\Gamma >0.3$, stable
breathing pulses do not exist. We stress that, unlike the effectively stable
nonbreathing pulses considered in section 2, we have {\em not} observed
stable breathing pulses on top of the small-amplitude background above the
zero-background's instability threshold in the region $\Gamma <0.3$ .

Bound states of stable localized pulses also play an important role in the
dynamics of GL-type models (see, e.g., \cite{Kawahara2} and \cite{Atai3}).
Our simulations have revealed a possibility of forming a stable bound state
of two breathers. Figure 12 displays a bound state found at $\gamma _{0}=0.18
$ and $\Gamma =0.2$. A noteworthy feature of this bound state is that the
phase shift between internal vibrations in the two breathers is $\pi /2$,
i.e., one breather has a maximum amplitude when the amplitude of its mate is
minimal, and vice versa. Attempts to generate a bound state of three
breathers have failed. In this connection, it is relevant to mention that
fully stable bound states of two stationary pulses can be readily generated
in the same model, while complexes consisting of three pulses exist, but
they are destroyed by perturbations
breaking their symmetry \cite{Atai3}.

\section{Conclusion}

In this work, we have considered a model based on the cubic complex
Ginzburg-Landau equation which is linearly coupled to an additional linear
equation, which was introduced in \cite{Atai1} as a simplification of a more
general model originally proposed in \cite{Winful}. The model describes a
dual-core nonlinear optical fiber with one active and one passive core. By
means of systematic simulations, we have found new types of stable localized
excitations in this model, existing in addition to the earlier found
stationary pulses. The newly found excitations include pulses existing on
top of a small-amplitude background (which is regular or chaotic) above the
threshold of instability of the zero solution, and breathers into which
stationary pulses are transformed by a Hopf bifurcation below the
zero-solution instability threshold. A sharp border between the stable
stationary pulses and breathers was identified. Stable bound states of two
breathers with a phase shift $\pi /2$ between their internal vibrations have
been found too. Above the threshold, the pulses are, respectively, standing
if the background oscillations are regular, and randomly walking if the
background is chaotic. With the increase of the system's size, additional
randomly walking pulses are spontaneously generated. The random walk in a
multi-pulse state is synchronized (but not completely) due to repulsion
between the pulses. At a large overcriticality, the multi-pulse state goes
over into a turbulent one. Lastly, breathers do not exist on top of the
small-amplitude background.

\smallskip One of the authors (B.A.M.) appreciates financial support from
the Japan Society for Promotion of Science, and hospitality of the
Department of Applied Physics at the Faculty of Engineering, Kyushu
University (Fukuoka, Japan).

\newpage

\section*{FIGURE CAPTIONS}

Fig. 1. The neutral stability curve $\gamma _{0}(q)$ at $\sigma =1$, $\gamma
_{1}=0.9$, $\gamma _{2}=5$, and $\Gamma =1.35$.

Fig. 2. The minimum value of the gain giving rise to instability of the zero
background vs. $\Gamma $ at $\sigma =1$, $\gamma _{1}=0.9$, and $\gamma
_{2}=5$.

Fig. 3. A stable solitary pulse existing on top of the small-amplitude
background at $\gamma _{0}=0.54$ and $\Gamma =1.35$: the wave field $|u(x)|$
(a), and the time dependence of ${\rm Re}\,u$ at two fixed spatial points,
one at the center of the pulse and one in the background (b).

Fig. 4. Snapshot of the wave field $|u(x,t)|$ in the localized pulse
existing on top of the chaotic background in the case $\gamma _{0}=0.61$ and 
$\Gamma =1.35$ (a), and the time evolution of $|u(x,t)|$ (b).

Fig. 5. Details of the random walk of the coordinate $X_{p}(t)$ (a), and
diffusive growth of the mean-square displacement $\left\langle \left(
X_{p}(t+\Delta t)-X_{p}(t)\right) ^{2}\right\rangle $ (b) of the solitary
pulse from fig. 4.

Fig. 6. Instantaneous snapshot (a) and the time evolution (b) of the wave
field $|u|$ at $\gamma _{0}=0.61$ and $\Gamma =1.35$. Synchronization of the
random walk of three spontaneously formed pulses is seen.

Fig. 7. The same as in fig. 6(b), but for the system four times as long. 
As an initial condition, we use four copies of the snapshot pattern 
of fig. 6(a), that is, 
$u(x,0)$ for $0<x<120$ takes the randomly perturbed value of fig. 6(a), and 
$u(x,0)$ for $120<x<240$ also takes the value of fig. 6(a) perturbed randomly, and so on. So initially there are 12 pulses.  
It is clearly seen that the synchronization of the random walk of different
pulses is imperfect.

Fig. 8. A ``turbulent'' configuration of $|u(x,t)|$ at $\gamma =1.05$ (a)
and $\gamma =2$ (b) for $\Gamma =1.35$.

Fig. 9. A stable breather observed at $\gamma _{0}=0.18$ and $\Gamma =0.2$:
the field $|u(x,t)|$ (a), and its value at the central point of the breather
vs. time (b).

Fig. 10. The amplitude $A$ of the oscillations of the breather's field at its
central point (see fig. 9b) vs. the gain parameter $\gamma _{0}$ (a), and $A^2$ vs. $\gamma_{0}$ near the critical point (b). The value $%
\gamma _{0}\approx 0.156$, at which the oscillation amplitude vanishes, is a
point of the transition (Hopf bifurcation) between the stationary pulse and
breather.

Fig. 11. The phase diagram in the parametric plane ($\Gamma $,$\gamma _{0}$%
), the other parameters taking the fixed values (\ref{fixed}). The zero
background is stable beneath the line marked by squares; localized pulses
disappear (decaying to zero) beneath the line marked by diamonds; crosses
mark the border between the stable stationary solitary pulses (above the
border) and stable breathers (below the border).

Fig. 12. A stable bound state of two breathers in the same case as in fig. 9.
Two pulses are set up apart from each other as an initial condition.
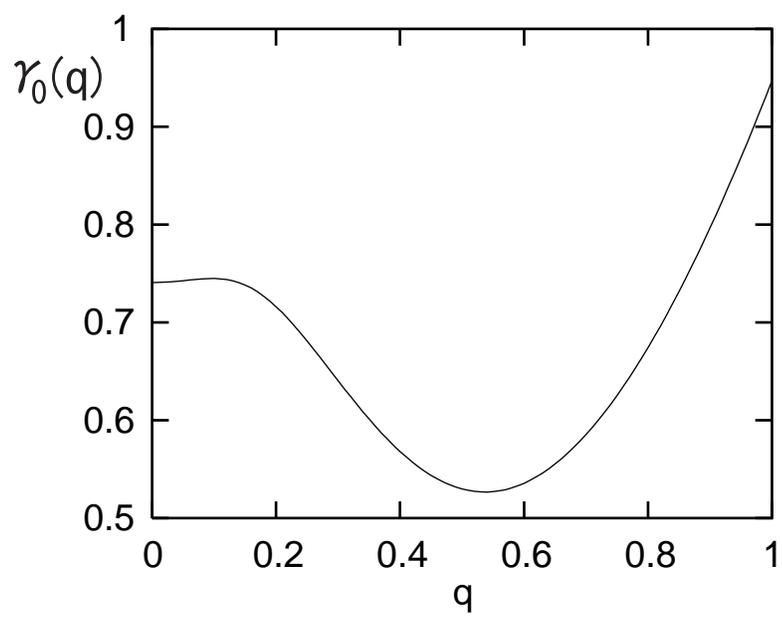
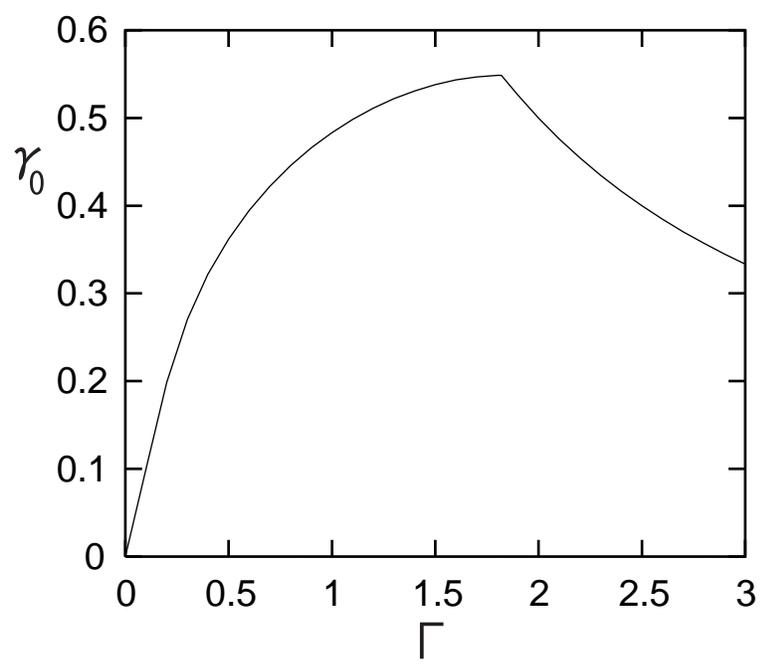
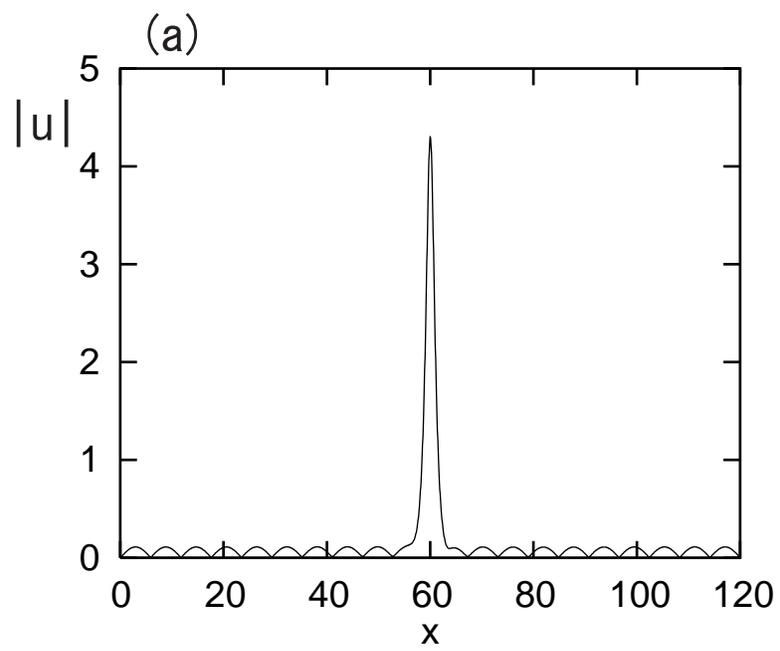
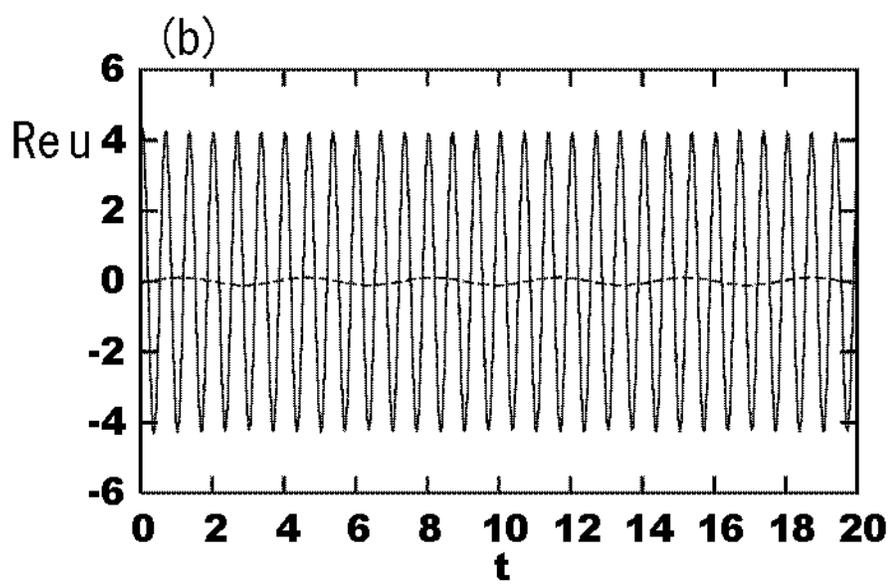
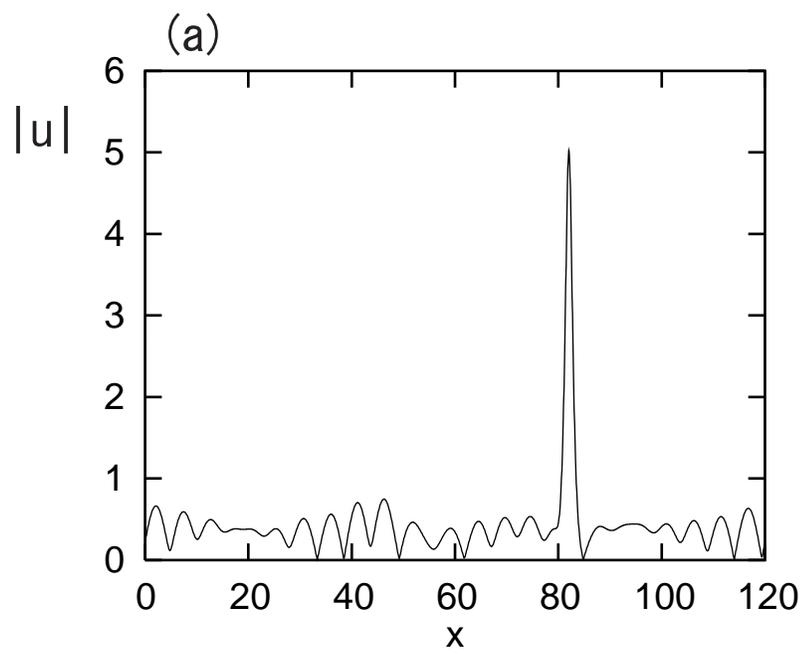
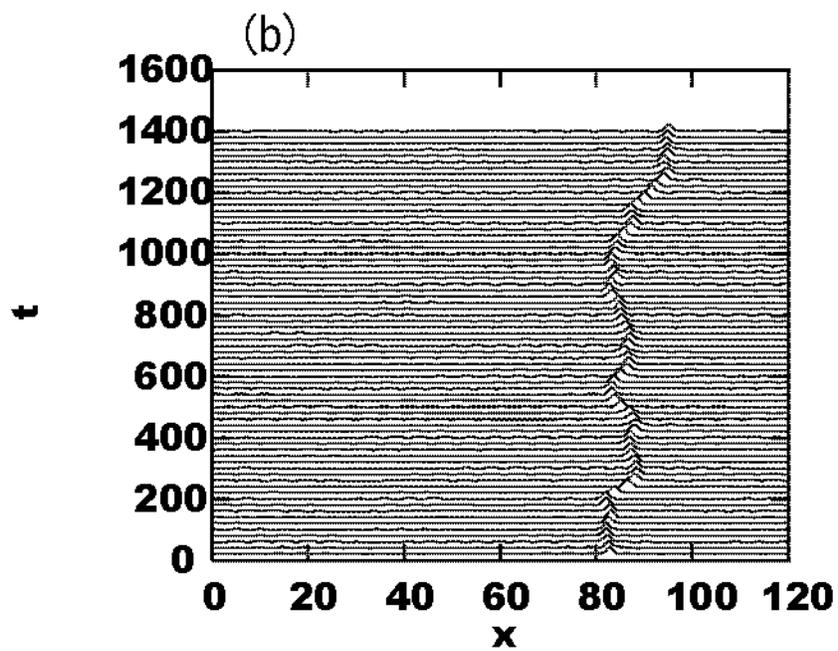
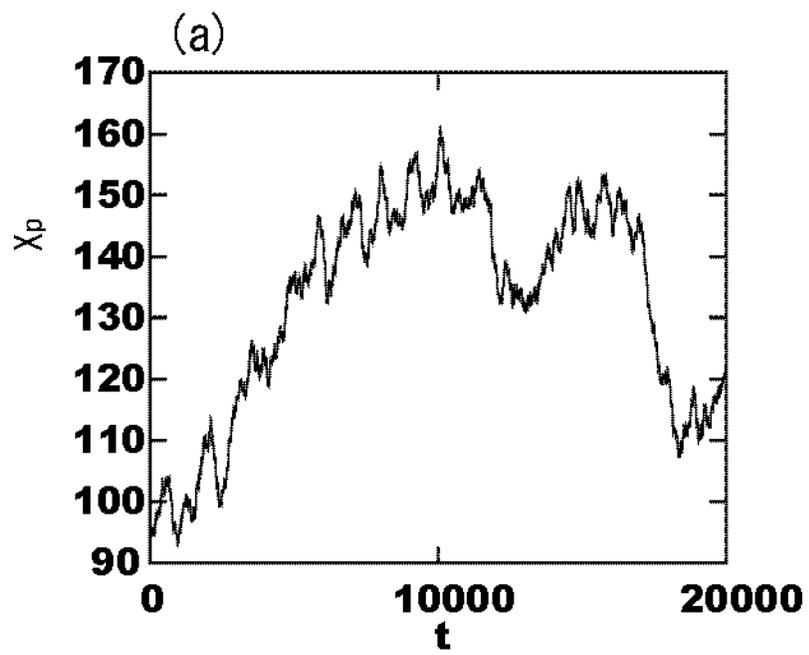
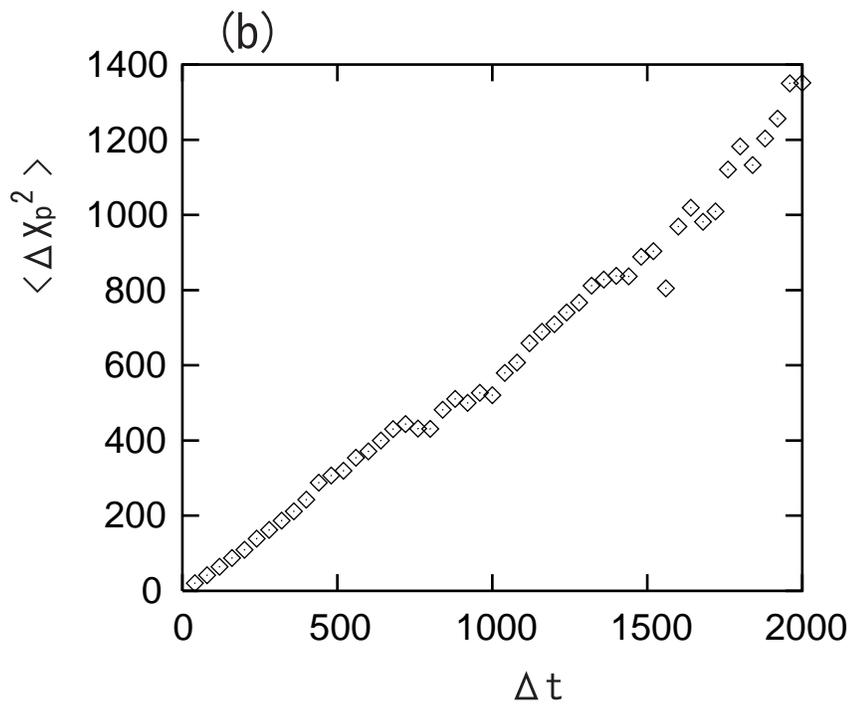
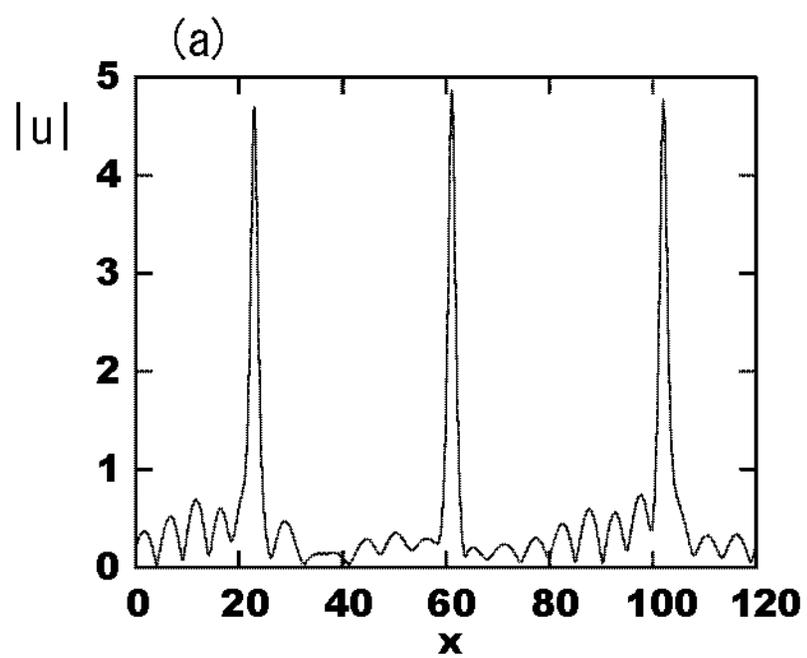
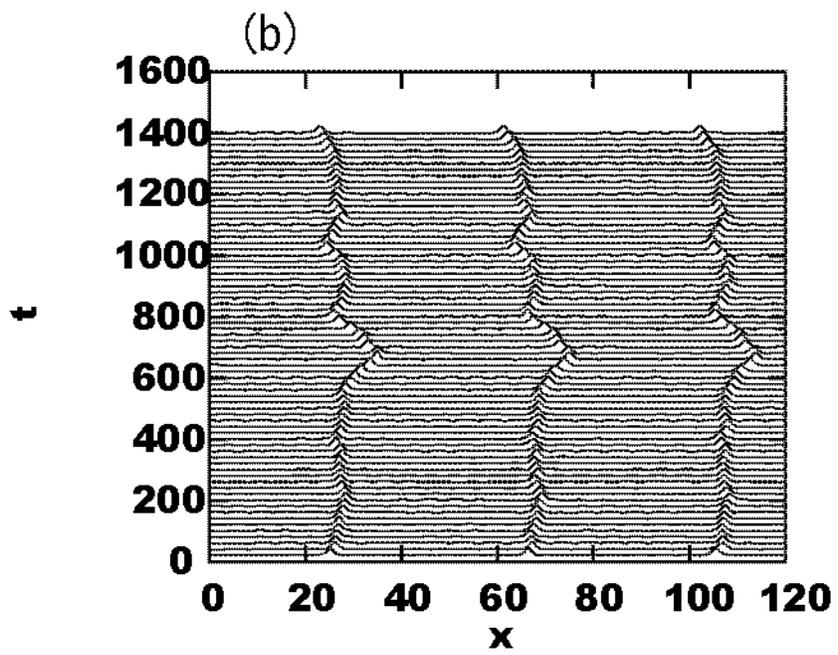
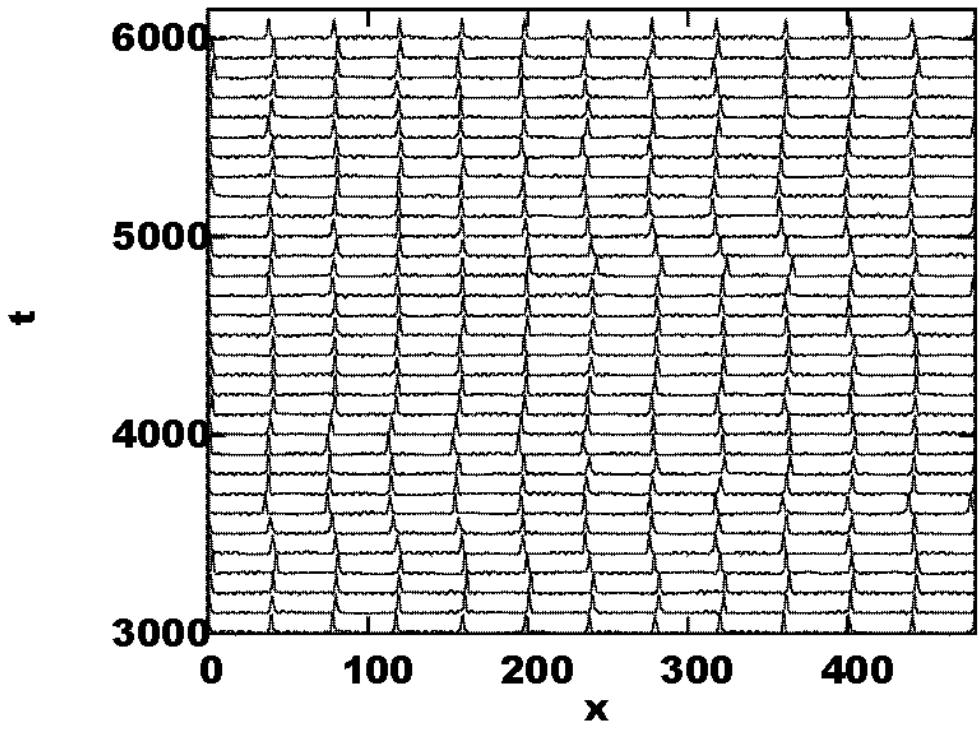
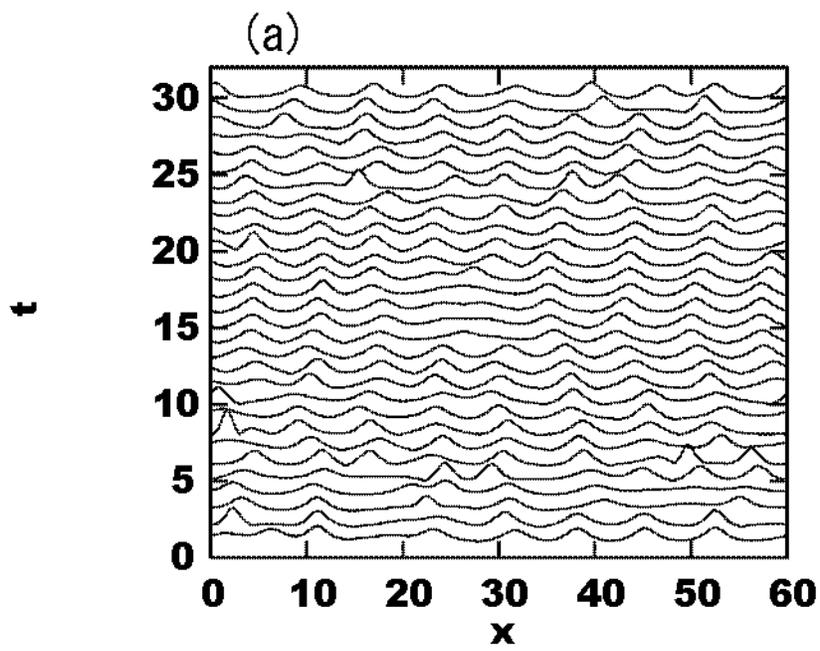
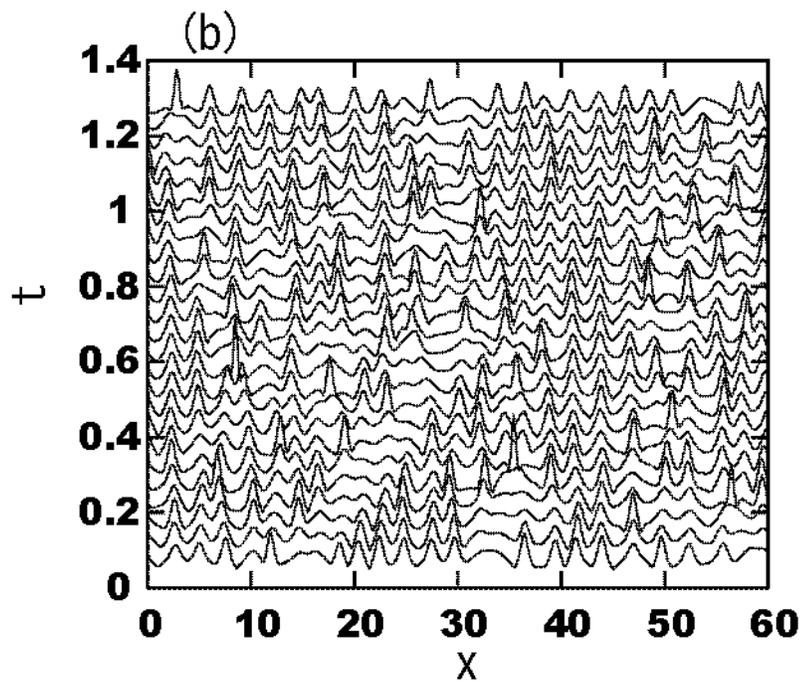
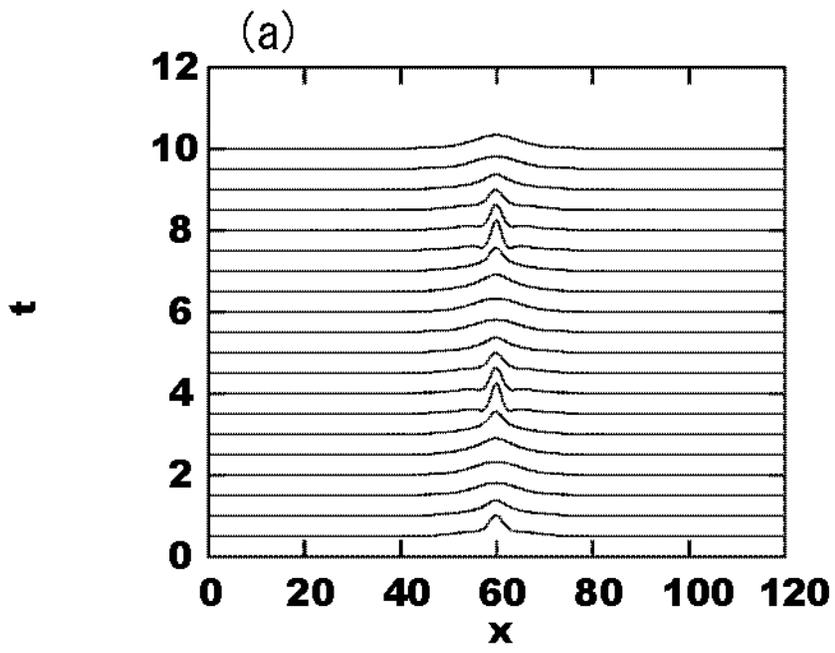
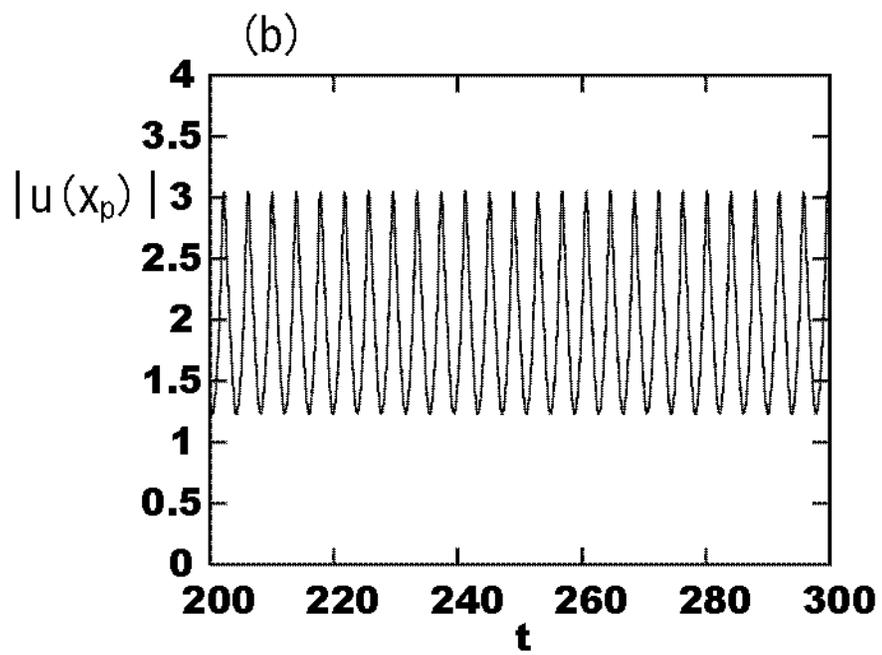
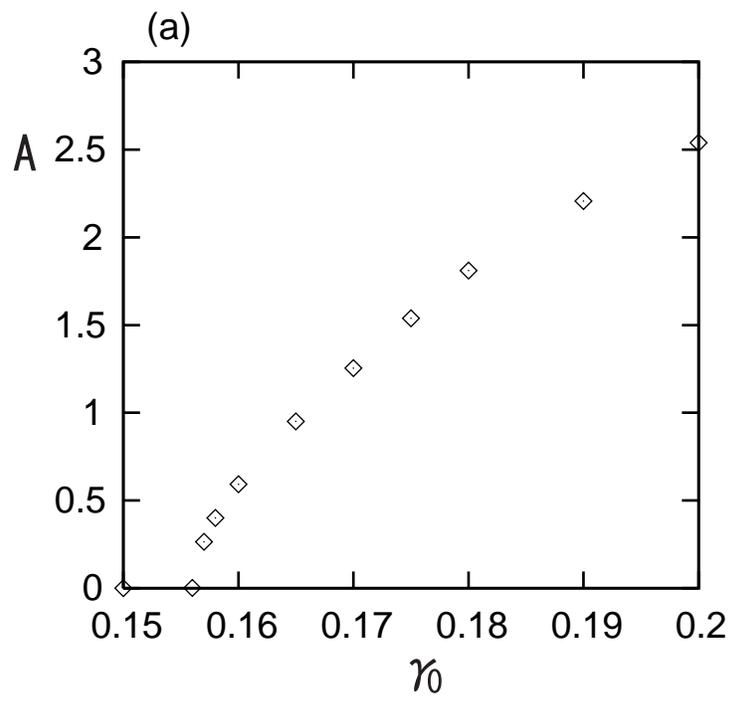
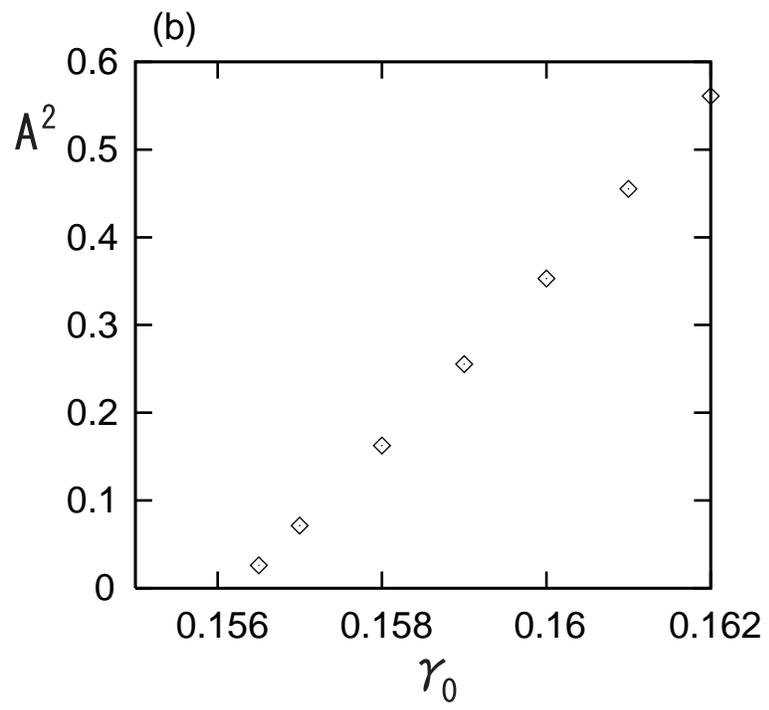
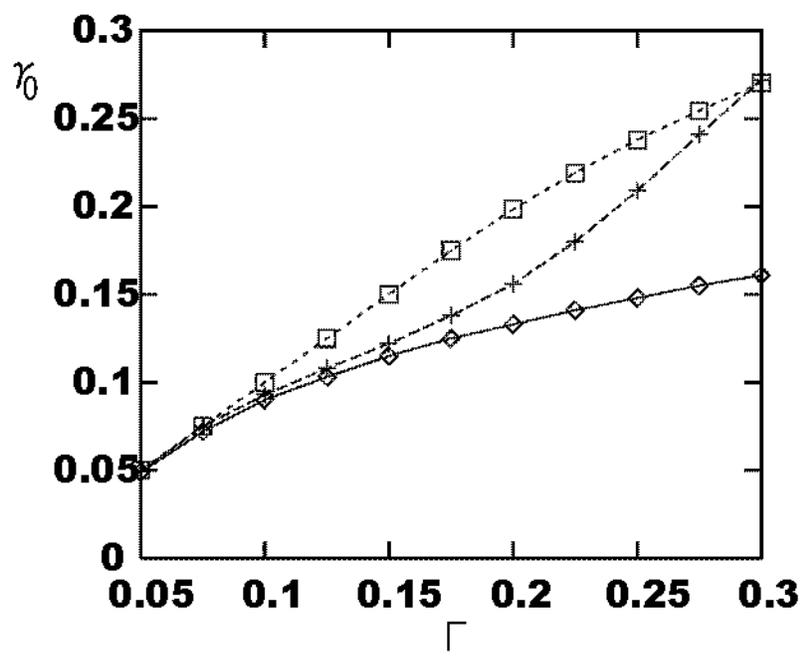
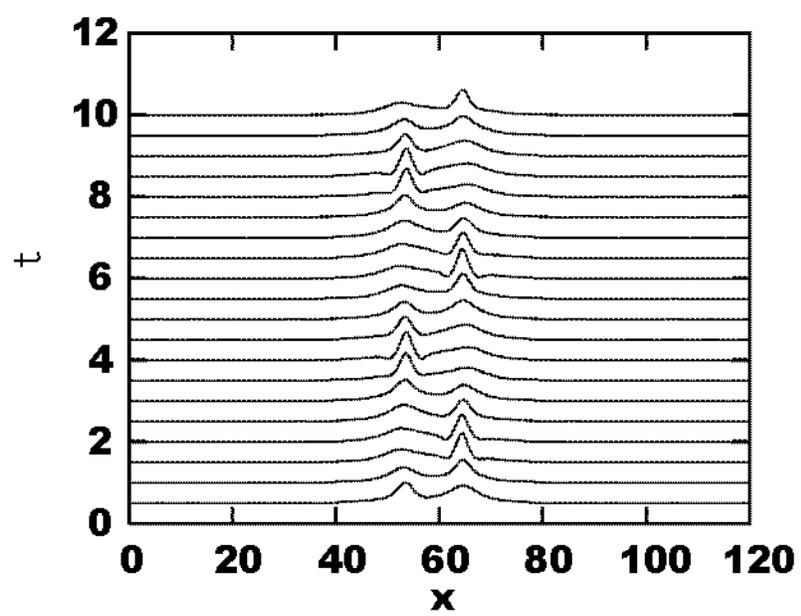
\end{document}